\documentclass[a4paper,12pt,final,twoside]{article}

\usepackage{amssymb,amsmath}

\usepackage[pdfstartview=FitH,pdfpagemode=None,bookmarks=true]{hyperref}

\usepackage{graphicx}
\usepackage{booktabs}

\newcommand{\figref}[1]{Fig.~\ref{#1}}

\DeclareMathAlphabet{\mathpzc}{OT1}{pzc}{m}{it}

\newcommand{\F}[2]{F^#1_{#2}}
\newcommand{\FA}{F_A}
\newcommand{\FP}{F_P}
\newcommand{\GF}{G_F}
\newcommand{\MA}{M_A}
\newcommand{\MV}{M_V}
\newcommand{\gA}{g_A}
\newcommand{\GE}{G^V_E}
\newcommand{\GM}{G^V_M}
\newcommand{\RA}{\mathcal R_{A}}
\newcommand{\RV}{\mathcal R_{V}}

\begin{document}

\title{High-energy limit of neutrino\\ quasielastic cross section}
\author{Artur M. Ankowski\footnote{{\scriptsize IFT University of Wroc{\l}aw, pl. M. Borna 9, 50-204 Wroc{\l}aw, Poland,~\href{mailto:artank@ift.uni.wroc.pl}
{artank@ift.uni.wroc.pl}}} }

\maketitle

\begin{abstract}
It's a common knowledge that the quasielastic neutrino-neutron and
anti\-neutrino-proton cross sections tend to the same constant as
(anti)neutrino energy becomes high. In this paper we calculate the
exact expression of the limit in terms of the parameters
describing quasielastic scattering. We check that even at very
high energies only small absolute values of the four-momentum
transfer contribute to the cross section, hence the Fermi theory
can be applied. The dipole approximation of the form factors
allows to perform analytic calculations. Obtained results are
neutrino-flavour independent.
\end{abstract}

\section{Introduction}

Quasielastic neutrino scattering plays a dominant role in
neutrino-nucleon reactions at energies below 1~GeV. When neutrino
energy increases another channels open and quasi\-elastic
processes become less important. At high energy the total cross
section for neutrino scattering is approximately proportional to
the value of the energy while the quasielastic cross section is
roughly constant. The latter behaviour is known on the basis of
numerical computations but as far as we know, it has not been
shown analytically yet.

The quasielastic cross section is usually calculated within the
Fermi theory. At low energies four-momentum transfer is understood
to fulfil the condition $|q^2|\ll M_W^2$, where $M_W=80{.}4$ GeV
is $W$-boson mass. It will be shown in Sec.~\ref{Discussion} that
in fact even for very-high-energy neutrinos overwhelming
contribution to the cross section satisfies such constraint,
therefore the use of the Fermi theory is well justified.

Radiative corrections are not taken into account, but in
Sec.~\ref{Discussion} we estimate that they can be neglected.

In the theoretical description of the neutrino-nucleon interaction
the ha\-dro\-nic current is expressed in terms of the four form
factors due to Lorentz invariance and assumption that there are no
second-class currents. The form factors can be expressed in
various ways, see~\cite{BBA}. We consider dipole form factors
because of their simplicity in analytic calculations.

The quasielastic cross section for neutrino-neutron scattering can
be written as~\cite{Llewellyn}
\begin{equation}\label{sigma_Llewellyn}
\sigma=\frac{M^2\GF^2\cos^2\theta_C}{8\pi E_\nu^2}\int\!d{q^2}\:\Big[A(q^2)-B(q^2)\frac{(s-u)}{M^2}+C(q^2)\frac{(s-u)^2}{M^4}\Big],
\end{equation}
where~$M=(m_n+m_p)/2$ is the average nucleon mass and
\[\begin{split}
A(q^2)&=\frac{m_l^2-q^2}{4M^2}\thinspace\Big[|\FA|^2\thinspace\Big(4-\frac{q^2}{M^2}\Big)-|\F{1}{V}|^2\Big(4+\frac{q^2}{M^2}\Big)\\
&\phantom{=\frac{m_l^2-q^2}{4M^2}\thinspace\Big[}-\frac{q^2}{M^2}|\xi\F{2}{V}|^2\thinspace\Big(1+\frac{q^2}{4M^2}\Big)-\frac{4q^2}{M^2}\Re\big(\F{1}{V}(\xi \F{2}{V})^*\big)\\
&\phantom{=\frac{m_l^2-q^2}{4M^2}\thinspace\Big[}-\frac{m_l^2}{M^2}\thinspace\Big(|\F{1}{V}+\xi\F{2}{V}|^2+|\FA|^2+4\Re(\FA\FP^*)+ \frac{q^2}{M^{2}}|\FP|^2\Big)\Big],\\
B(q^2)&=-\frac{q^2}{M^2}\Re\big((\F{1}{V}+\xi \F{2}{V})\FA^*\big),\\
C(q^2)&=\frac1 4\Big(|\F{1}{V}|^2-\frac{q^2}{4M^2}|{\xi\F{2}{V}}|^2+|\FA|^2\Big).
\end{split}\]
In above formulae $m_l$ is charged-lepton mass, $E_\nu$ ---
neutrino energy and $\xi=\mu_p-\mu_n-1$, where $\mu_p$ and $\mu_n$
are the proton and neutron magnetic moments respectively. In the
case of antineutrino-proton scattering $-B(q^2)$ in
Eq.~\eqref{sigma_Llewellyn} should be replaced by $+B(q^2)$. We
also need to know the interval of integration $\big[(q^2)_A,
(q^2)_B\big]$:
\begin{equation}\label{interval}
\begin{split}
(q^2)_A&=\frac{m_l^2(E_\nu+M)-2ME_\nu^2-\sqrt{\varDelta}}{2E_\nu+M},\\
(q^2)_B&=\frac{m_l^4M}{m_l^2(E_\nu+M)-2ME_\nu^2-\sqrt{\varDelta}},
\end{split}
\end{equation}
with $\varDelta=(2ME_\nu^2-m_l^2E_\nu)^2-4m_l^2M^2E_\nu^2$.

As it was mentioned before, in this paper we will consider dipole
form factors. Using the Sachs form factors
\begin{align*}
\GE(q^2)&=\frac{1}{(1-q^2/\MV^2)^2},&\GM(q^2)=\frac{1+\xi}{(1-q^2/\MV^2)^2},
\end{align*}
the vector form factors can be expressed in the following way:
\[\begin{split}
\F{1}{V}(q^2)&=\Big({1-\frac{q^2}{4M^2}}\Big)^{-1}\Big[{\GE(q^2)-\frac{q^2}{4M^2}\GM(q^2)}\Big],\\
\xi\F{2}{V}(q^2)&=\Big({1-\frac{q^2}{4M^2}}\Big)^{-1}\Big[{-\GE(q^2)+\GM(q^2)}\Big],
\end{split}\]
whereas the pseudoscalar form factor~$\FP$ is related to the axial
one due to PCAC hypothesis:
\begin{align*}
&\FA(q^2)=\frac\gA{(1-q^2/\MA^2)^2},
&\FP(q^2)=\frac{2M^2\FA(q^2)}{m_\pi^2-q^2}.
\end{align*}
By~$m_\pi$ we denoted the pion mass.

\begin{table}
\centering
\begin{tabular}{lrr}
\toprule
$\GF$&1.1803&$10^{-5}/$GeV$^{2}$\\
$\cos\theta_C$&0.9740&\\
$\gA$&-1.267&\\
$\xi$&3.7059&$\mu_N$\\
$\MA$&1.001&GeV\\
$\MV^2$&0.71&GeV$^2$\\
\bottomrule
\end{tabular}
\vspace{9pt} \caption{The values of the constants used in
numerical calculations}\label{constants} \vspace{-7pt}
\end{table}

\section{High-energy limit}
If neutrino energy~$E_\nu$ is high enough to fulfil the
condition~$M^\text{max}/E_\nu\ll1$,
where~$M^\text{max}=\max\{m_l,M,\MV,\MA\}$, one can write
\[
{\varDelta}=4M^2E_\nu^2\Big[E_\nu^2-\frac{m_l^2}{M}E_\nu+\frac{m_l^4}{4M^2}-m_l^2\Big]\rightarrow 4M^2E_\nu^4,
\]
what results in
\begin{equation}\label{HE_interval}
\begin{split}
(q^2)_A&\rightarrow-2ME_\nu,\\
(q^2)_B&\rightarrow0.
\end{split}
\end{equation}
The cross section~\eqref{sigma_Llewellyn} is the sum of terms
\begin{equation}\label{alpha, beta, kappa}
\begin{split}
\alpha&\doteq\frac{\mathcal G}{4E_\nu^2}\int\!d{q^2}M^2A(q^2),\\
-\beta&\doteq\frac{\mathcal G}{4E_\nu^2}\int\!d{q^2}B(q^2)\,(s-u),\\
\kappa&\doteq\frac{\mathcal G}{4E_\nu^2}\thinspace\int\!d{q^2}C(q^2)\frac{(s-u)^2}{M^2},
\end{split}
\end{equation}
where we have introduced the compact notation for the constant
factor
\[
\mathcal G=\frac{\GF^2\cos^2\theta_C}{2\pi}.
\]

The first term, that is~$\alpha$, tends to zero as neutrino
energy becomes infinite. We will show it in
Appendix~\ref{Proof:A}.

Next, in Appendix~\ref{Proof:B} it is calculated directly that in
the discussed limit~$\beta$ also approaches zero. This result is
consistent with Pomeranchuk's theorem (see generalization
in~\cite{Weinberg}), which states that as~$E_\nu\rightarrow\infty$
the neutrino and antineutrino cross sections become equal. They
differ in sign of the $\beta$~term, so the term should tend to
zero.

\begin{figure}
\centering
\includegraphics*[bb=115 308 470 522, scale=0.97]{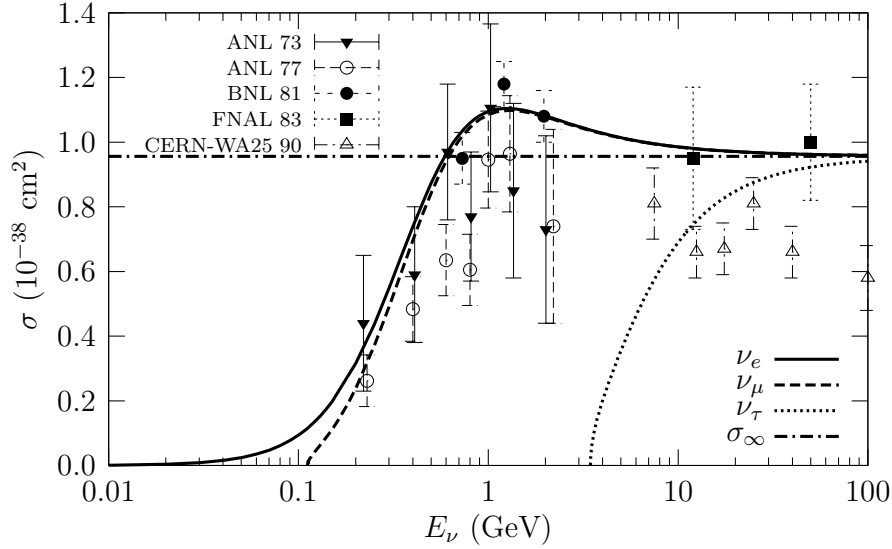}
\caption{The cross sections' dependence on neutrino energy.
$\sigma_\infty$~stands for the high-energy limit of~$\sigma$
calculated in this paper. Experimental data for quasielastic
$\nu_\mu$ scattering from $D_2$ target are taken from
ANL~1973~\protect\cite{73Mann}, ANL~1977~\protect\cite{77Barish},
BNL~1981~\protect\cite{81Baker},
FNAL~1983~\protect\cite{83Kitagaki} and
CERN-WA25~1990~\protect\cite{90Allasia}.}\label{figCS3f}
\end{figure}

Thus only $\kappa$ gives a nonzero contribution to the high-energy
(anti)neu\-trino quasi\-elastic cross section:
\[
\sigma_\infty\doteq\lim_{E_\nu\rightarrow\infty}\sigma=\lim_{E_\nu\rightarrow\infty}\kappa.
\]
Our main result can be can be written in the form
\[\begin{split}
\sigma_\infty=&\frac{\GF^2\cos^2\theta_C}{6\pi}\Big[{\MV^2}+{\gA^2\MA^2}+\frac{2\xi(\xi+2)\MV^4}{(4M^2-\MV^2)^2}(M^2-\MV^2)\\
&\mspace{188mu}+\frac{3\xi(\xi+2)\MV^8}{(4M^2-\MV^2)^3}\Big(\frac{4M^2}{4M^2-\MV^2}\ln\frac{4M^2}{\MV^2}-1\Big)\Big].
\end{split}\]
Detailed calculations are presented in Appendix~\ref{Proof:C}.
Introducing notation~$\rho=4M^2/\MV^2$ and~$\mu=\xi+1$ we can
write it in the more compact way:
\begin{equation}\label{limit_compact}
\begin{split}
\sigma_\infty=&\frac{\GF^2\cos^2\theta_C}{6\pi}\Big[{\MV^2}+{\gA^2\MA^2}+2M^2\frac{\mu^2-1}{(\rho-1)^2}\,\Big(1-\frac4{\rho}\Big)\\
&\mspace{188mu}+3\MV^2\frac{\mu^2-1}{(\rho-1)^3}\Big(\frac{\rho}{\rho-1}\ln(\rho)-1\Big)\Big].
\end{split}
\end{equation}
The above expression does not depend on the charged-lepton mass,
therefore the $E_\nu\rightarrow\infty$ limit of the cross section
is equal for all the neutrinos and antineutrinos, see
also~\figref{figCS3f}.

\begin{figure}
\centering
\includegraphics*[bb=105 308 460 522, scale=0.97]{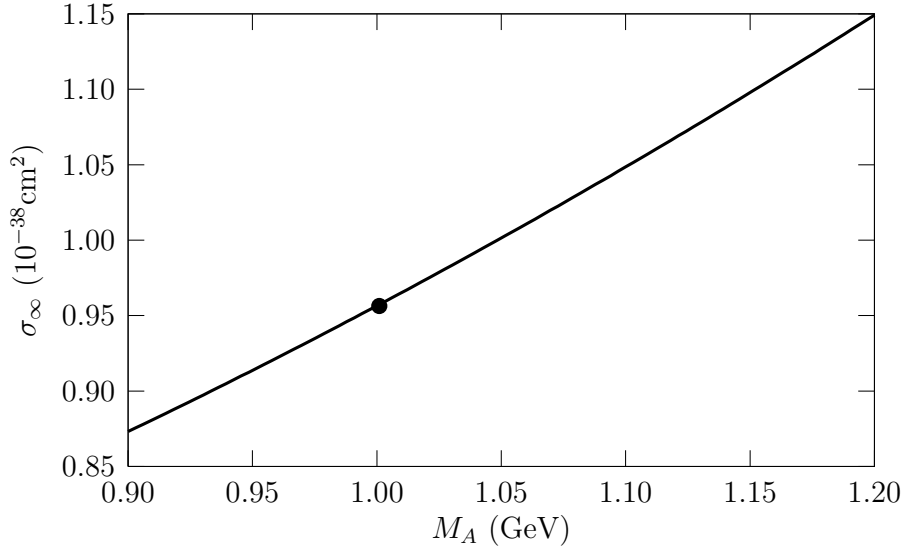}
\caption{The dependence of the high-energy limit of the cross
section~$\sigma_\infty$ on the axial mass. Marked point represents
the value of $\MA$ as in~\protect\cite{BBA}.}\label{figMA}
\end{figure}

\section{Discussion}\label{Discussion}
To obtain numerical value of the limit we assume values of the
constants for dipole form factors as in~\cite{BBA}, see
Tab.~\ref{constants}. Note the corrected value of the axial mass:
$\MA=1{.}001\pm0.020$ GeV. Then
\[
\sigma_\infty=0{.}956\times10^{-38}\text{ cm}^2.
\]
We observe next that none of the four terms in
Eq.~\eqref{limit_compact} can be neglected. Contribution of the
term with the axial form factor is equal to about 46\%. The
dependence of~$\sigma_\infty$ on the value of the axial mass is
shown in Fig.~\ref{figMA}.

\begin{figure}
\centering
\includegraphics*[bb=115 308 470 522, scale=0.97]{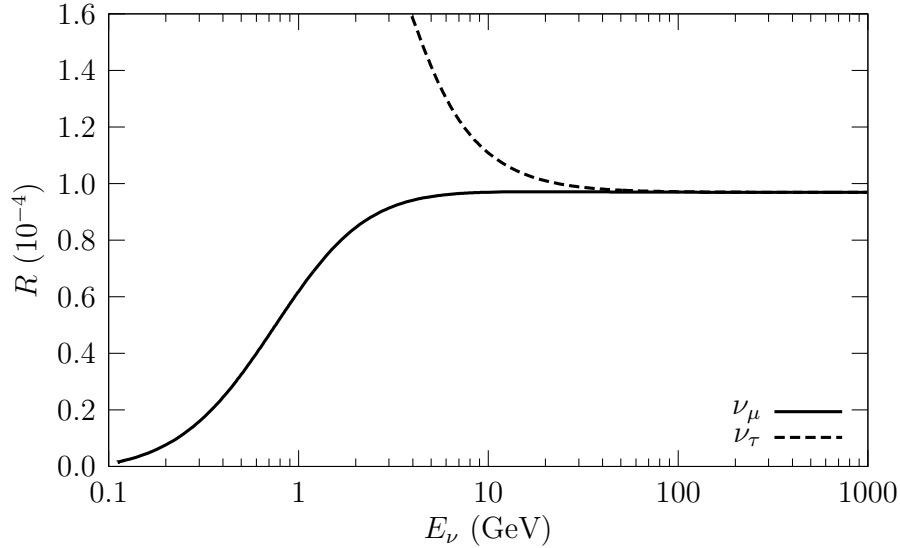}
\caption{The ratio of the difference between the cross section~$\sigma_W$
with the $W$-boson propagator and the Fermi-theory cross section normalized
with respect to~$\sigma_W$ itself.}\label{figR}
\end{figure}

It is necessary to check if our approach based on the Fermi theory
is consistent. We do it numerically by computing the cross section
with the $W$-boson propagator~$\sigma_W$ and comparing the result
with the cross section within the Fermi theory~$\sigma$.
\figref{figR} presents the dependence of the ratio
\[
R=\frac{\sigma_W-\sigma}{\sigma_W}
\]
on neutrino energy. When $E_\nu\geq50$ GeV the ratio~$R$ is
roughly constant and less than~0.01\% (for each flavour). It means
that only small four-momentum transfers~$|q^2|$ contribute to the
quasielastic cross section, thus calculations within the Fermi
theory are reasonable even for very high neutrino energies.

In calculations of the limit of the cross section no radiative
corrections were taken into account. We guess that corrections to
quasielastic scattering are of the same order of magnitude as to
deep inelastic scattering, i.e. they are roughly constant and of
the order of half a percent~\cite{Sigl} (the value refers to the
corrections which comes from  bremsstrahlung of the charged
lepton, $W$~boson and quarks). If the hypothesis is true, it makes
them of low importance unless experiments reach very high
precision.

More important improvements could come from the non-dipole form
factors as in~\cite{BBA}. Presented there figures suggest that they
would yield the value of the limit 3\% smaller with respect
to our result, but unfortunately ``BBA-2003 Form Factors'' are
practically unapplicable to analytic calculations.

\appendix
\section[Appendix A]{Why $\alpha$ tends to zero}\label{Proof:A}
The $\alpha$ term defined in Eq.~\eqref{alpha, beta, kappa} is an
integral of rational function of~$q^2$ divided by neutrino energy
squared. As~$E_\nu\rightarrow\infty$, $\alpha$ wouldn't tend to
zero only if the in\-te\-gral rose at least as~$E_\nu^2$. The form
of the limits~\eqref{HE_interval} implies that the lower one
always gives zero and only the upper one could produce nonzero
terms, if the integrand is of the order at least one in~$q^2$.
Let's write explicitly the term of the highest order for each form
factor, keeping in memory that~$\FP$ can be expressed by~$\FA$:
\[\begin{split}
\frac14\Big(\frac{q^2}{M^2}\Big)^2|\FA|^2&=\frac{\gA^2\MA^8}{4M^4}\frac{(q^2)^2}{(\MA^2-q^2)^4},\\
\frac14\Big(\frac{q^2}{M^2}\Big)^2|\F{1}{V}|^2&=\frac{\MV^8}{4M^4}\frac{\big(q^2\big)^2}{\big(4M^2-q^2\big)^2}\frac{\big(4M^2-q^2(\xi\!+\!1)\big)^2}{\big(\MV^2-q^2\big)^4},\\
\frac1{16}\Big(\frac{q^2}{M^2}\Big)^3|\xi\F{2}{V}|^2&=\frac{\xi^2\MV^8}{M^2}\frac1{(4M^2-q^2)^2}\frac{(q^2)^3}{(\MV^2-q^2)^4}.
\end{split}\]
We can see that each one of them is a~proper fraction, so as
neutrino energy becomes infinite $\alpha$ tends to zero. To be
sure, let's perform the calculation for the second of above
expressions. We can obtain easy-to-integrate form by decomposing
it into partial fractions:
\[\begin{split}
\frac14\Big(\frac{q^2}{M^2}\Big)^2|\F{1}{V}|^2&=\frac{\MV^8}{4M^4}\Big[\frac{c}{(4M^2-q^2)}-\frac{c}{(\MV^2-q^2)}+\mathcal O(1/q^2)\Big]
\end{split}\]
where~$c$ is a~constant and~$\mathcal O(1/q^2)$ denotes terms of
lower order in~$q^2$. As neutrino energy becomes high the limits
of integration are given by~\eqref{HE_interval} hence
\[\begin{split}
\frac14\int\!d{q^2}\Big(\frac{q^2}{M^2}\Big)^2|\F{1}{V}|^2&\rightarrow\frac{\MV^8}{4M^4}\Big[{2c}\ln\frac{\MV}{2M}+\text{other constants}\Big].
\end{split}\]
The above integral tends to a~constant as
$E_\nu\rightarrow\infty$. Only higher order term in~$q^2$ could
give result increasing with~$E_\nu$ but there is no such term
in~$\alpha$. Since that for the whole expression holds true that
\[
\alpha\rightarrow\frac{\text{const}}{E_\nu^2}\rightarrow0.
\]

\section[Appendix B]{Why $\beta$ tends to zero}\label{Proof:B}
In the frame in which target nucleon is at
rest~$(s-u)=(4ME_\nu-m_l^2+q^2)$, so the quantity defined in
Eq.~\eqref{alpha, beta, kappa} can be explicitly written as
\[
\beta=\frac{\mu\gA \mathcal G(\MA\MV)^4}{4M^2E_\nu^2}\int\!d{q^2}\big(\mathpzc B(4ME_\nu-m_l^2)+\mathpzc B q^2\big),
\]
where
\[
\mathpzc B=\frac{q^2}{(\MA^2-q^2)^2(\MV^2-q^2)^2}.
\]
To perform the integration one need to decompose the integrand
into partial fractions:
\[\begin{split}
\mathpzc B&=\frac1{\RA^2}\Big[\frac{\MA^2}{(\MA^2-q^2)^2}+\frac{\MV^2}{(\MV^2-q^2)^2}+\frac{\MA^2\!+\!\MV^2}{\RA}\Big(\frac1{\MA^2-q^2}-\frac1{\MV^2-q^2}\Big)\Big],\\
\mathpzc Bq^2&=\frac1{\RA^2}\Big[\frac{\MA^4}{(\MA^2-q^2)^2}+\frac{\MV^4}{(\MV^2-q^2)^2}+\frac{2\MA^2\MV^2}{\RA}\Big(\frac{1}{\MA^2-q^2}-\frac{1}{\MV^2-q^2}\Big)\Big],
\end{split}\]
where~$\RA=\MA^2-\MV^2$. As~$M^\text{max}/E_\nu\ll1$, after
integrating in the limits~\eqref{HE_interval} we obtain
\[
\beta\rightarrow\frac{2\mu\gA \mathcal G(\MA\MV)^4}{ME_\nu\RA^2}\Big(1+\frac{\MA^2+\MV^2}{\RA}\ln\frac{\MV}{\MA}\Big)\rightarrow 0.
\]

\section[Appendix C]{Why $\kappa$ tends to constant}\label{Proof:C}
The last term in Eq.~\eqref{alpha, beta, kappa} expressed by the
form factors is
\[
\kappa=\frac{\mathcal G}{(4ME_\nu)^2}\int\!d{q^2}\Big[|\F{1}{V}|^2-\frac{q^2}{4M^2}|{\xi\F{2}{V}}|^2+|\FA|^2\Big]{(s-u)^2}.
\]
For convenience we separate the axial part from the vector one:
\[\begin{split}
\kappa_\text{\tiny A}&\doteq\frac{\mathcal G}{(4ME_\nu)^2}\int\!d{q^2}\thinspace|\FA|^2(s-u)^2,\\
\kappa_\text{\tiny V}&\doteq\frac{\mathcal G}{(4ME_\nu)^2}\int\!d{q^2}\Big[|\F{1}{V}|^2-\frac{q^2}{4M^2}|\xi\F{2}{V}|^2\Big](s-u)^2.
\end{split}\]
To evaluate the integral
\[
\kappa_\text{\tiny A}=\frac{\gA^2\mathcal G
\MA^8}{(4ME_\nu)^2}\int\!d{q^2}\frac{(s-u)^2}{(\MA^2-q^2)^4}.
\]
one needs to know decomposition of the integrand. If we add~$\MA$
and~$-\MA$ to $(s-u)$ and square it in the following way
\[
(s-u)^2=(4\!M\!E_\nu-m_l^2+\MA^2)^2-2(4\!M\!E_\nu-m_l^2+\MA^2)(\MA^2-q^2)+(\MA^2-q^2)^2,
\]
we will get
\[
\frac{(s-u)^2}{(\MA^2-q^2)^4}=\frac{(4\!M\!E_\nu-m_l^2+\MA^2)^2}{(\MA^2-q^2)^4}-\frac{2(4\!M\!E_\nu-m_l^2+\MA^2)}{(\MA^2-q^2)^3}+\frac1{(\MA^2\!-\!q^2)^2}.
\]
It means that as neutrino energy fulfils
condition~$M^\text{max}/E_\nu\ll1$, integration in the
limits~\eqref{HE_interval} leads to
\[
\kappa_\text{\tiny A}\rightarrow\frac{\gA^2\mathcal G\MA^2}{3}\bigg(1-\frac{2m_l^2+\MA^2}{4ME_\nu}\bigg)
\]
and
\[
\lim_{E_\nu\rightarrow\infty}\kappa_\text{\tiny A}=\mathcal G\,\frac{\gA^2\MA^2}{3}.
\]
The integrand in definition of~$\kappa_\text{\tiny V}$, i.e.
\[\begin{split}
&|\F{1}{V}|^2-\frac{q^2}{4M^2}|\xi\F{2}{V}|^2={\Big(\!1\!-\!\frac{q^2}{4M^2}\!\Big)^{-1}\!\Big(1\!-\!\frac{q^2}{\MV^2}\Big)^{-4}}\Big[\mu^2\Big(1\!-\!\frac{q^2}{4M^2}\Big)\!+\!1\!-\!\mu^2\Big]
\end{split}\]
with~$\mu=\xi+1$, can be written as
\[
|\F{1}{V}|^2-\frac{q^2}{4M^2}|\xi\F{2}{V}|^2=\frac{\mu^2\MV^8}{(\MV^2-q^2)^4}-\frac{4M^2\MV^8(\mu^2-1)}{(4M^2-q^2)(\MV^2-q^2)^4}.
\]
Let's denote the last-fraction's numerator as $\mathcal
K=4M^2\MV^8(\mu^2-1)$. Above expression decomposed into partial
fractions is
\[\begin{split}
|\F{1}{V}|^2-\frac{q^2}{4M^2}|\xi\F{2}{V}|^2&=\frac{\mathcal K}{\RV^4}\bigg(\frac1{\MV^2-q^2}-\frac1{4M^2-q^2}\bigg)+\frac{\mathcal K}{\RV^3(\MV^2-q^2)^2}\\
&\mspace{156mu}+\frac{\mathcal K}{\RV^2(\MV^2-q^2)^3}+\frac{\mathcal K_\mu}{\RV(\MV^2-q^2)^4},
\end{split}\]
where $\RV=4M^2-\MV^2$ and $\mathcal
K_\mu=\MV^8(4M^2-\mu^2\MV^2)$. By repeating the trick made during
the computation of~$\kappa_\text{\tiny A}$ we obtain
\[\begin{split}
\Big[|\F{1}{V}|^2-\frac{q^2}{4M^2}|\xi\F{2}{V}|^2\Big](s-u)^2&=\frac{c_1}{\MV^2-q^2}-\frac{c_1}{4M^2-q^2}+\frac{c_2}{(\MV^2-q^2)^2}\\
&\mspace{129mu}+\frac{c_3}{(\MV^2-q^2)^3}+\frac{c_4}{(\MV^2-q^2)^4},
\end{split}\]
where coefficients are:
\[\begin{split}
c_1&=\frac{\mathcal K}{\RV^4}(4ME_\nu-m_l^2+4M^2)^2,\\
c_2&=\frac{\mathcal K}{\RV^3}\Big[\frac{\mu^2\MV^8\RV^3}{\mathcal K}-(4ME_\nu-m_l^2+4M^2)^2\Big],\\
c_3&=\frac{1}{\RV^2}\Big[\mathcal K\Big(4ME_\nu-m_l^2+\MV^2-\frac{\mathcal K_\mu\RV}{\mathcal K}\Big)^2-\frac{(\mathcal K_\mu\RV)^2}{\mathcal K}\Big],\\
c_4&=\frac{\mathcal K_\mu}{\RV}(4ME_\nu-m_l^2+\MV^2)^2.
\end{split}\]
For neutrino energy~$E_\nu\gg M^\text{max}$, we conclude that
integration over~$(d q^2)$ leads to
\[\begin{split}
&\frac{\mathcal G}{(4ME_\nu)^2}\int\!d{q^2}\Big(\mspace{-2mu}\frac{c_1}{\MV^2\mspace{-2mu}-\mspace{-2mu}q^2}-\frac{c_1}{4M^2\mspace{-2mu}-\mspace{-2mu}q^2}\mspace{-2mu}\Big)\mspace{-2mu}\rightarrow\mspace{-2mu}\frac{\mathcal{GK}}{\RV^4}\ln\frac{4M^2}{\MV^2}\Big(1+\frac{4M^2-m_l^2}{2ME_\nu}\Big),\\
&\frac{\mathcal G}{(4ME_\nu)^2}\int\!d{q^2}\frac{c_2}{(\MV^2-q^2)^2}\rightarrow-\frac{\mathcal{G K}}{\MV^2\RV^3}\Big(1+\frac{4M^2-m_l^2}{2ME_\nu}\Big),\\
&\frac{\mathcal G}{(4ME_\nu)^2}\int\!d{q^2}\frac{c_3}{(\MV^2-q^2)^3}\rightarrow\frac{\mathcal G}{2\MV^4\RV^2}\Big(\mathcal K+\frac{\mathcal K(\MV^2-m_l^2)-\mathcal K_\mu\RV}{2ME_\nu}\Big),\\
&\frac{\mathcal G}{(4ME_\nu)^2}\int\!d{q^2}\frac{c_4}{(\MV^2-q^2)^4}\rightarrow\frac{\mathcal G\mathcal K_\mu}{3\MV^6\RV}\Big(1+\frac{\MV^2-m_l^2}{2ME_\nu}\Big).
\end{split}\]
The $\kappa$ term is the sum of~$\kappa_\text{\tiny A}$
and~$\kappa_\text{\tiny V}$, therefore
\[
\lim_{E_\nu\rightarrow\infty}\kappa=\mathcal G\,\frac{\gA^2\MA^2}{3}+\frac{\mathcal G}{3\RV}\Big[\frac{3\mathcal K}{\RV^3}\ln\frac{4M^2}{\MV^2}-\frac{3\mathcal K}{\MV^2\RV^2}+\frac{3\mathcal K}{2\MV^4\RV}+\frac{\mathcal K_\mu}{\MV^6}\Big].
\]
Recall that~$\mathcal K_\mu=\MV^8(4M^2-\mu^2\MV^2)$
and~$\RV=4M^2-\MV^2$, hence we obtain
\[\begin{split}
\frac{\mathcal K_\mu}{\MV^6}&=\MV^2\big(4M^2-\mu^2\MV^2\big)=\MV^2\RV-(\mu^2-1)\MV^4.
\end{split}\]
Next, constant factor $\mathcal K=4M^2\MV^8(\mu^2-1)$, so
\[
\frac{\mathcal K_\mu}{\MV^6}+\frac{3\mathcal K}{2\MV^4\RV}=\MV^2\RV+(\mu^2-1)\MV^4\frac{2(M^2-\MV^2)+3\MV^2}{\RV}.
\]
It means that the limit of the cross section is equal to
\[\begin{split}
\lim_{E_\nu\rightarrow\infty}\sigma&=\frac{\GF^2\cos^2\theta_C}{6\pi}\Big[{\MV^2}+{\gA^2\MA^2}+\frac{2(\mu^2-1)\MV^4}{(4M^2-\MV^2)^2}(M^2-\MV^2)\\
&\mspace{184mu}+\frac{3(\mu^2-1)\MV^8}{(4M^2-\MV^2)^3}\Big(\frac{4M^2}{4M^2-\MV^2}\ln\frac{4M^2}{\MV^2}-1\Big)\Big].
\end{split}\]
Denoting~$\rho=4M^2/\MV^2$ we can write this formula in the
following way:
\[\begin{split}
\lim_{E_\nu\rightarrow\infty}\sigma=&\frac{\GF^2\cos^2\theta_C}{6\pi}\Big[{\MV^2}+{\gA^2\MA^2}+2M^2\frac{\mu^2-1}{(\rho-1)^2}\,\Big(1-\frac4{\rho}\Big)\\
&\mspace{234mu}+3\MV^2\frac{\mu^2-1}{(\rho-1)^3}\Big(\frac{\rho}{\rho-1}\ln(\rho)-1\Big)\Big].
\end{split}\]

\end{document}